\documentclass[preprint2,twocolumn,times,tighten]{aastex631}
\usepackage{subfigure}
\usepackage{newtxtext,newtxmath}
\usepackage{amsmath}
\usepackage{hyperref}
\usepackage{amssymb}
\usepackage{natbib}
\usepackage{morefloats}
\usepackage{multirow}
\usepackage{array}
\usepackage{verbatim}
\usepackage{color}

\let\tablenum\relax
\usepackage{siunitx}

\usepackage{graphicx}	
\usepackage{amsmath}	
\newcommand*{\yh}[1]{\textcolor{red}{#1}}

\begin{document}

\title{Machine-Learning Estimation of Energy Fractions in MHD Turbulence Modes}

\author{Jiyao Zhang}
\affiliation{Department of Mathematics, University of Pennsylvania, Philadelphia, PA 19104, USA}

\author[0000-0002-8455-0805]{Yue Hu*}
\affiliation{Institute for Advanced Study, 1 Einstein Drive, Princeton, NJ 08540, USA}


\email{jiyaoz@sas.upenn.edu; yuehu@ias.edu; *NASA Hubble Fellow}



\begin{abstract}
Magnetohydrodynamic (MHD) turbulence plays a central role in many astrophysical processes in the interstellar medium (ISM), including star formation and cosmic-ray transport and acceleration. MHD turbulence can be decomposed into three fundamental modes—fast, slow, and Alfv\'en—each contributing differently to the dynamics of the medium. However, characterizing and separating the energy fractions of these modes was challenging due to the limited 2D information available from observations. To address this difficulty, we use 3D isothermal and multiphase MHD turbulence simulations to examine how mode energy fractions vary under different physical conditions. Overall, we find that the Alfv\'en and slow modes carry comparable kinetic-energy fractions and together dominate the turbulent energy budget in multiphase media, while the fast mode contributes the smallest fraction. Relative to isothermal conditions, multiphase simulations exhibit an enhanced fast-mode energy fraction. We further introduce a machine–learning–based approach that employs a conditional Residual Neural Network to infer these fractions directly from spectroscopic data. The method leverages the fact that the three MHD modes imprint distinct morphological signatures in spectroscopic maps owing to their differing contributions to density and velocity fluctuations. Our model is trained on a suite of isothermal and multiphase simulations covering typical ISM conditions. We demonstrate that our machine learning model can recover the mode fractions from spectroscopic observables, achieving mean relative normalized errors of approximately 0 and standard deviation of 0.01 - 0.02 for seen data and 0.1 - 1.8 for unseen data.
\end{abstract}



\section{Introduction} \label{sec:intro}

Magnetohydrodynamic (MHD) turbulence is a fundamental component of the interstellar medium (ISM), shaping its dynamical, thermal, and chemical evolution \citep{1998JGR...103.1889L, 1995ApJ...438..763G,2003PhPl...10.1954N, 2010ApJ...708.1204B,2016ApJ...832..143F,2020MNRAS.498.1593B,2021ApJ...907L..40H,2024A&A...691A.303T, 2025ApJ...986...62H,2025ApJ...983...32H,2025MNRAS.544.4173N}. Through the cascade of energy from large to small scales, it governs the structure and dynamics of the multiphase ISM and thereby regulates star formation \citep{2004RvMP...76..125M, 2012nsf....1211729M,2012ApJ...761..156F,2023MNRAS.524.4431H}. MHD turbulence mediates the transport of heat, momentum, and mass between different ISM phases \citep{2009SSRv..143..333D,2025ApJ...986...62H} and strongly influences the propagation and acceleration of cosmic rays \citep{1966ApJ...146..480J, 2002PhRvL..89B1102Y, 2008ApJ...673..942Y, 2013ApJ...779..140X,2022MNRAS.514..657K,2023FrASS..1054760L,2025arXiv250507421H,2025arXiv250907104H,2026MNRAS.545f2108E}. These processes collectively establish MHD turbulence as a central agent in linking microphysical plasma processes to the large-scale evolution of galaxies.

Over the past few decades, theoretical advances have greatly deepened our understanding of MHD turbulence \citep{1995ApJ...438..763G, LV99}, while numerical simulations have provided crucial tests of these theoretical frameworks \citep{2000ApJ...539..273C}. In particular, \citet{CL03} analyzed the scaling and anisotropy of compressible MHD turbulence by decomposing it into three fundamental eigenmodes: fast, slow, and Alfv\'en modes. The Alfv\'en and slow modes follow a Kolmogorov-like $k^{-5/3}$ spectrum with the scale-dependent anisotropy \citep{LV99}, whereas fast modes exhibit an isotropic $k^{-3/2}$ spectrum in subsonic condition or $k^{-2}$ spectrum in supersonic condition \citep{CL03,2010ApJ...720..742K, 2022MNRAS.512.2111H,2025ApJ...992L..28H}.

Isotropic fast modes play a dominant role in scattering cosmic rays through gyro-resonant interactions, whereas the highly anisotropic Alfv\'en and slow modes are less efficient due to the averaging effects over many eddies during gyro-resonance \citep{2002PhRvL..89B1102Y, 2004ApJ...614..757Y,2022MNRAS.512.2111H,2023FrASS..1054760L}. Consequently, the relative energy fraction distribution among the three modes is a key factor in determining cosmic-ray transport in the ISM, yet constraining this distribution observationally remains highly challenging.

In this work, we propose a machine-learning approach to estimate the energy partition among Alfv\'en, slow, and fast modes in the ISM using spectroscopic observations. This approach leverages the fact that these three modes imprint distinct morphological signatures in spectroscopic maps due to their differing anisotropy and compressibilities. These characteristic differences leave identifiable imprints on spectroscopic data. Machine-learning methods, such as Convolutional Neural Networks (CNNs; \citealt{lecun1998gradient}), which excel at extracting such morphological patterns, therefore offer a promising avenue for predicting the energy partition among the three modes and constraining the mode composition of MHD turbulence in the ISM.

This paper is organized as follows. \S~\ref{sec:theory} presents the theoretical framework for mode decomposition and explains how morphological features—intensity, centroid velocity, and channel maps—encode information about MHD turbulence modes in both isothermal and multi-phase ISM conditions. \S~\ref{sec:data} describes the MHD turbulence simulations employed in this study and outlines the ResNet-based neural network architecture and training methodology. \S~\ref{sec:results} presents the results of our numerical experiments. \S~\ref{sec:dis} compares our findings with previous work and discusses future prospects of the machine-learning approach. Finally, \S~\ref{sec:con} summarizes the key results and conclusions.

\section{Theoretical consideration}
\label{sec:theory}

\subsection{Properties of MHD modes}
\begin{figure*}[h!]
    \centering
    \includegraphics[width=1.0\linewidth]{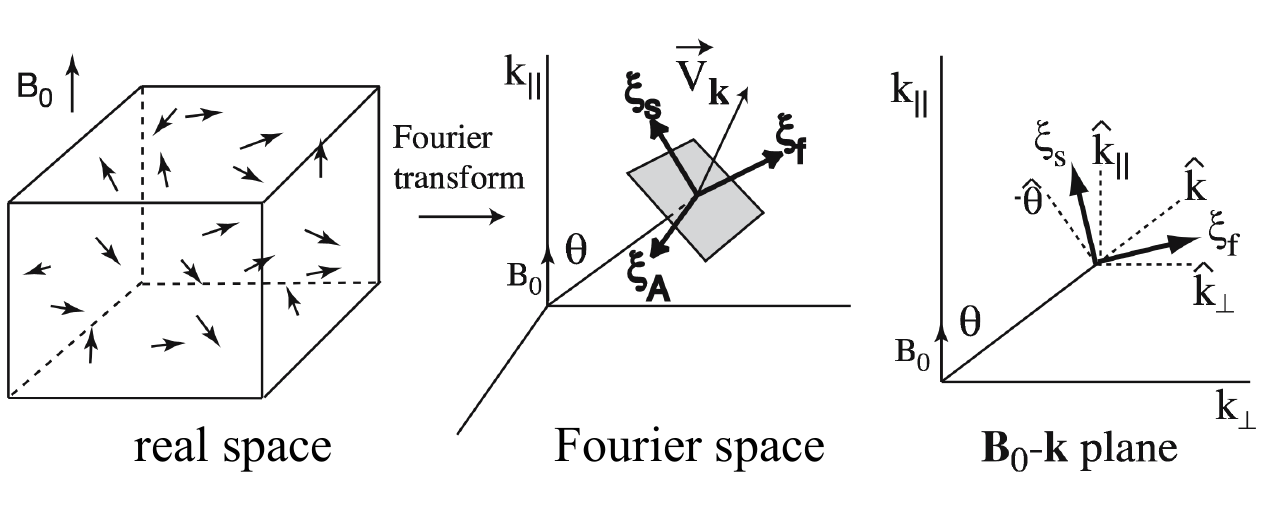}
    \caption{Mode decomposition method. We separate Alfv\'en, slow, and fast modes in Fourier space by projecting the velocity Fourier component $\pmb{v_k}$ onto bases $\hat{\xi}_{\rm a}$, $\hat{\xi}_{\rm s}$, and $\hat{\xi}_{\rm f}$, respectively. Note that $\hat{\xi}_{\rm s}$ and $\hat{\xi}_{\rm f}$ lie in the plane defined by $\mathbf{B_0}$ and $\hat{\mathbf{k}}$. Slow basis $\hat{\xi}_{\rm s}$ lies between $-\hat{\theta}$ and $\hat{\mathbf{k}}_\parallel$. Fast basis $\hat{\xi}_{\rm f}$ lies between $\hat{\mathbf{k}}$ and $\hat{\mathbf{k}}_\bot$. From \cite{2003MNRAS.345..325C}. }
    \label{fig:mode_method}
\end{figure*}

In ideal compressible MHD turbulence, fluctuations can be decomposed into three distinct modes: the incompressible Alfv\'en mode and two compressible modes, commonly referred to as the fast and slow modes. \citet{2002PhRvL..88x5001C, 2003MNRAS.345..325C} introduced a decomposition method to study these three modes. The corresponding basis vectors \(\hat{\boldsymbol{\xi}}\) defining the mode decomposition are given by:
\begin{equation}
\label{eq:modedecomposition}
\begin{aligned}
\hat{\boldsymbol{\xi}}_{\rm s} &\propto \left(1+\frac{\beta'}{2}-\sqrt{D}\right)k_{\bot}\hat{k}_{\bot} + \left(-1+\frac{\beta'}{2}-\sqrt{D}\right)k_{\parallel}\hat{k}_{\parallel},\\
\hat{\boldsymbol{\xi}}_{\rm f} &\propto \left(1+\frac{\beta'}{2}+\sqrt{D}\right)k_{\bot}\hat{k}_{\bot} + \left(-1+\frac{\beta'}{2}+\sqrt{D}\right)k_{\parallel}\hat{k}_{\parallel}, \\
\hat{\boldsymbol{\xi}}_{\rm a} &\propto \hat{k}_{\parallel} \times \hat{k}_{\bot},
\end{aligned}
\end{equation}
where \(D = (1+\beta'/2)^2 - 2\beta'\cos^2\theta\), \(\beta' = 2M_A/M_s\), and \(\theta\) is the angle between the wavevector \(\hat{\mathbf{k}}\) and the mean magnetic field \(\mathbf{B}_0\). Here the subscripts \(a\), \(f\), and \(s\) refer to the Alfv\'en, fast, and slow modes, respectively (see Fig.~\ref{fig:mode_method}). $M_A$ and $M_s$ are the Alfv\'en and sonic Mach numbers, respectively.

\begin{figure*}[htbp!]
    \centering
    \includegraphics[width=1.0\linewidth]{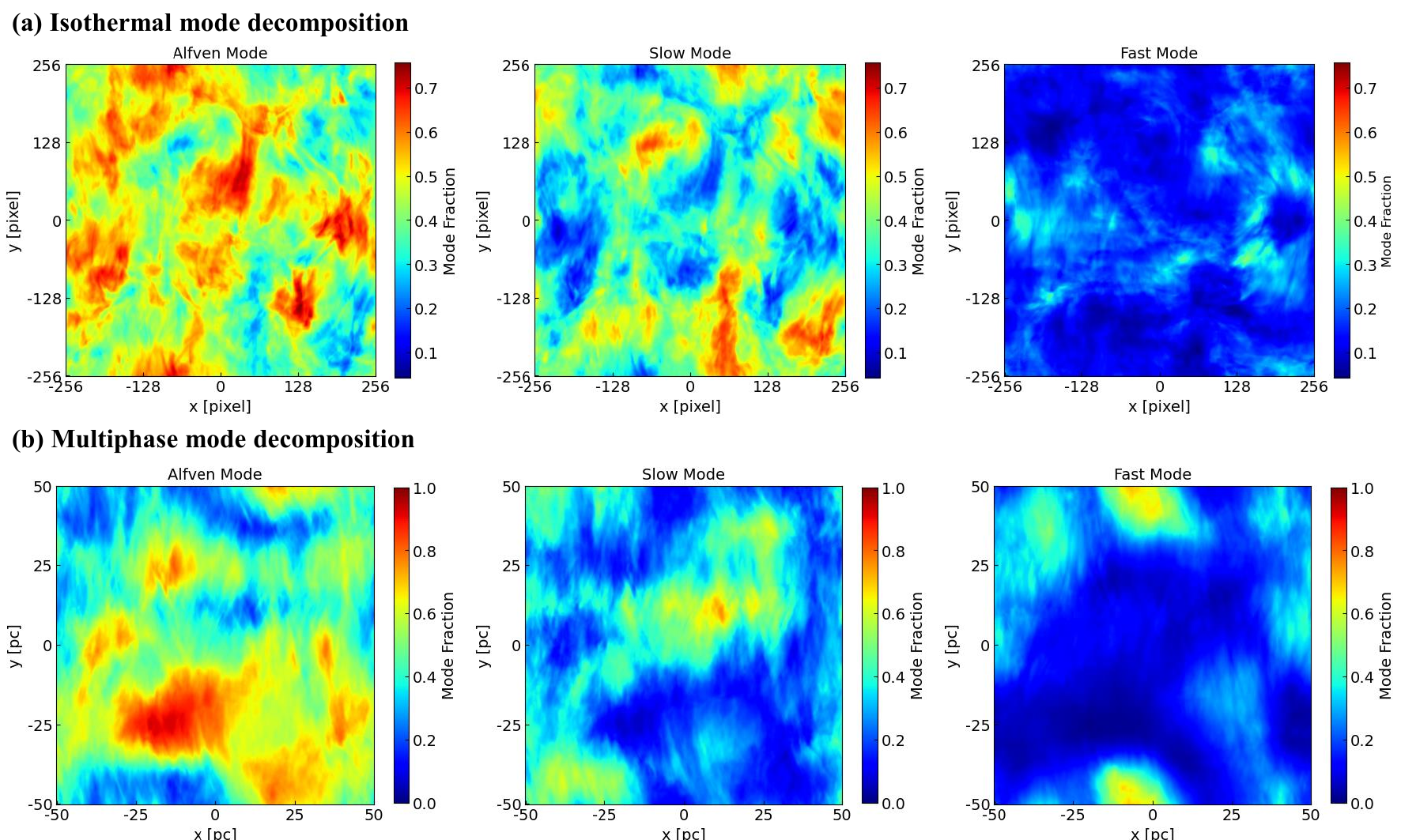}
    \caption{Illustration of three MHD turbulence modes' {\yh kinetic energy fraction} after mode decomposition. Panel (a): isothermal simulation min05 ($\beta = 0.02$). Panel (b): multi-phase simulation B6v5 ($\beta = 0.0061$).}
    \label{fig:mode_illustration}
\end{figure*}

Two representative examples of the mode decomposition for isothermal and multi-phase MHD turbulence are shown in Fig.~\ref{fig:mode_illustration}. The Alfv\'en mode is incompressible and propagates along the mean magnetic field, generating shear motions that stretch structures without producing appreciable density variations. The compressible slow mode behaves similarly to the pseudo-Alfv\'en mode in high-$\beta$ plasmas, whereas in low-$\beta$ environments it propagates at approximately the sound speed mainly along the magnetic field. As a result, the slow mode produces density and velocity fluctuations that are preferentially aligned with the field, giving rise to elongated, field-parallel filamentary structures. In contrast, the fast mode exhibits nearly isotropic propagation in high-$\beta$ plasmas, traveling at roughly the sound speed independent of field orientation and producing correspondingly isotropic, clumpy structures. In low-$\beta$ plasmas, the fast mode instead represents magnetic-pressure compressions that propagate at the Alfv\'en speed and predominantly compress the gas perpendicular to the magnetic field.

These distinct characteristics of the different modes imply that their relative energy fractions can produce markedly different features in velocity and density distributions, which can be probed through spectroscopic observations. Machine learning (ML), with its strength in recognizing and quantifying complex morphological patterns, offers a powerful approach to identify these mode-dependent structures and to predict the energy partition among MHD modes directly from observational data.

\begin{figure}
    \includegraphics[width=\columnwidth]{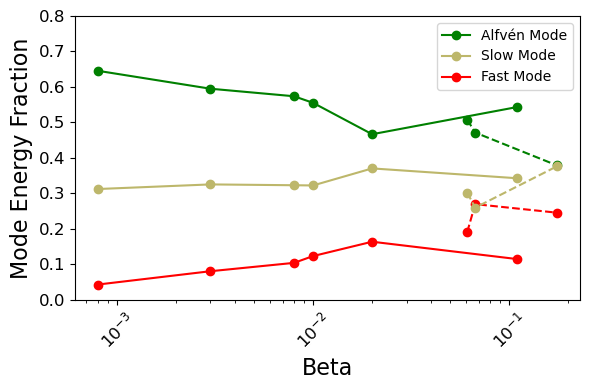}
    \caption{Dependence of energy fractions of the three MHD modes on plasma $\beta$. The colors represent different modes — red for Fast Mode, yellow for Slow Mode, and green for Alfv\'en Mode. Solid lines denote isothermal simulation, while dotted lines correspond to multi-phase simulation.}
    \label{beta_modeenergy}
\end{figure}

\section{Methodology} 
\label{sec:data}

\begin{figure*}[htbp!]
    \centering
    \includegraphics[width=1.0\linewidth]{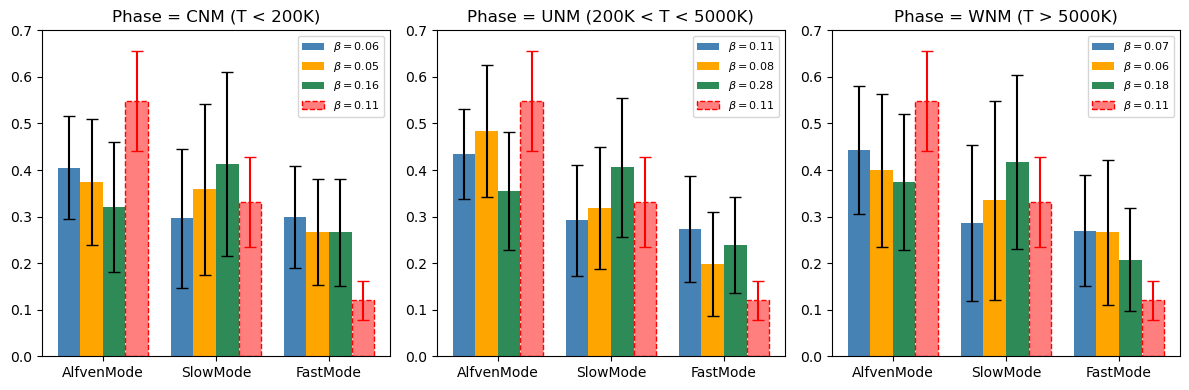}
    \caption{Boxplot of energy fraction of Alfv\'en mode, Slow mode, and Fast mode in three separate multi-phase simulations and one isothermal simulation. Error bars indicate 1 sigma variance. From left to right: CNM, UNM, and WNM phase. Color differs from simulation. The red box refers to the isothermal simulation, which is consistent between phases.}
    \label{fig:mode_precentage}
\end{figure*}

\subsection{3D isothermal MHD turbulence simulation}
The 3D isothermal MHD turbulence simulations analyzed in this study were generated using the ZEUS-MP/3D code. These simulations solve the ideal MHD equations with periodic boundary conditions, given by:
\begin{equation}
\label{eq.zeus}
\begin{aligned}
& \frac{\partial\rho}{\partial t} + \nabla \cdot (\rho \pmb{v})  = 0, \\
& \frac{\partial(\rho \pmb{v})}{\partial t} + \nabla \cdot \left[ \rho \pmb{v} \pmb{v}^T + (c_s^2\rho + \frac{B^2}{8\pi})I - \frac{\pmb{B}\pmb{B}^T}{4\pi} \right]  = \pmb{f},  \\
& \frac{\partial \pmb{B}}{\partial t} - \nabla \times (\pmb{v} \times \pmb{B}) = 0, \\
\end{aligned}
\end{equation}
where $\pmb{f}$ is a forcing term. We also consider a zero-divergence condition $\nabla \cdot \pmb{B} = 0$, and an isothermal equation of state $P = c_s^2 \rho_0$, where $P$ is the gas pressure, $\rho_0$ is the initial gas density, and $c_s$ is the sound speed. The initial conditions include a uniform density field and a uniform magnetic field aligned along the y-axis. 

\begin{table}[!ht]
\caption{Overview of isothermal simulation runs in this work. $\beta=2(M_A/M_s)^2$ is the plasma beta. $\delta\rho/\langle \rho\rangle=\sqrt{\langle(\rho-\langle \rho\rangle)^2\rangle}/\langle \rho\rangle$ is the density RMS fluctuation normalized to the uniform background density.}
\centering
\hspace*{-1.5cm}
\begin{tabular}{ccccc}
\hline\hline
\textbf{Model} & $M_s$ & $M_A$ & $\beta$ & $\delta\rho/\langle \rho\rangle$ \\ \hline
min01 & 11.05 & 0.22 & 0.0008 & 0.400 \\
min02 & 11.02 & 0.43 & 0.003 & 0.521 \\
min03 & 10.86 & 0.69 & 0.008 & 0.534\\
min04 & 10.50 & 0.81 & 0.010 & 0.568\\
min05 & 10.22 & 1.07 & 0.020 & 0.573 \\
min06 &  2.17 & 0.51 & 0.110 & 0.273 \\
\hline
\end{tabular}
\end{table}

Kinetic energy is solenoidally injected at wavenumber 2 and turbulence is naturally evolved to produce a Kolmogorov spectrum. We continuously drive turbulence and dump the data until the turbulence gets fully developed at one sound crossing time. The simulation is grid into $512^3$ cells. The simulations are characterized by the sonic Mach number $M_s$ and Alfv\'en Mach number $M_A$. In this work, we refer to the simulations in Table 1 by their model name or key parameters. 

\subsection{3D Multiphase ISM simulations}
The 3D multiphase MHD turbulence simulations analyzed were generated using the AthenaK code \citep{2024arXiv240916053S}. In addition to the ideal MHD equations, cooling and heating functions are included in the energy equation:
\begin{equation}
\label{eq.athenak}
\begin{aligned}
& \frac{\partial E}{\partial t} + \nabla \left[ v(E+P+\frac{\pmb{B}^2}{8\pi}) - \frac{\pmb{B}(\pmb {B}\cdot \pmb{v})}{4\pi} \right] = \Gamma - \Lambda
\end{aligned}
\end{equation}
where $\Gamma$ and $\Lambda$ are the heating energy density rate and cooling energy density rate given by \citep{2002ApJ...564L..97K}: 
\begin{equation} 
\begin{aligned} 
\Lambda = &\left(\frac{\rho}{m_{\rm H}}\right)^2\Bigl[2\times10^{-19}\exp\left(\frac{-114800}{T+1000}\right)\\ 
&+ 2.8\times10^{-28}\sqrt{T}\exp\left(\frac{-92}{T}\right)\Bigr] ~~ {\rm erg~s^{-1}cm^{-3}}, \\
\Gamma = & \left(\frac{\rho}{m_{\rm H}}\right)\times10^{-26} ~~ {\rm erg~s^{-1}~cm^{-3}}. 
\end{aligned} 
\end{equation} 
where $m_{\rm H}$ is the hydrogen mass. 

Similarly, the simulation box size is 100 pc, with turbulence driven solenoidally at a peak wavenumber of 2. An example of compressively driven turbulence is presented in the Appendix's Fig.~\ref{fig:mode_illustration_compressible}. The computational domain is discretized on a uniform $512^3$ cells. The initial conditions include a uniform number density field of $n = 3 \text{cm}^{-3}$ and a uniform magnetic field aligned along the y-axis. We choose $B \approx3~\SI{}{\micro G}$ and $B \approx 6~\SI{}{\micro G}$ based on the Zeeman observation \citep{crutcher2012magnetic}. Two values of velocity dispersion are included: $\sigma_v =2.5$ and $\sigma_v = 5~\text{km s}^{-1}$. These parameters are summarized in Table~2. For details, we refer to \cite{2025ApJ...986...62H}.

\begin{table}[!ht]
\caption{Overview of multi-phase simulation runs in this work. Here $\beta$ is calculated from the mean $M_A$ amd $M_s$.}
\centering
\hspace*{-1cm}
\resizebox{0.9\columnwidth}{!}{
\begin{tabular}{cccccc}
\hline\hline
\textbf{Model} & $B$ [\SI{}{\micro G}] & $\sigma_v$ [km/s] & $\beta$ & $\delta\rho/\langle \rho\rangle$ & \\ \hline
B6v2.5 & 6.0 & 2.5 & 0.067 & 0.379 \\
B3v2.5 & 3.0 & 2.5 & 0.175 & 0.551  \\
B6v5 & 6.0 & 5.0 & 0.061 & 0.451 \\
\hline
\end{tabular}}
\end{table}
 

\subsection{Synthetic spectroscopic observation}
To create synthetic spectroscopic observations as our ML training input, we utilize velocity and density fields derived from MHD simulations to construct PPV cubes. Following early work on $M_s$ prediction \citep{2025ApJ...982..121S}, we generate intensity maps, velocity centroids, and velocity channel maps as our model training morphological features.

Intensity maps, $I(x, y)$, are produced by fully integrating the PPV cubes along the LOS, which is defined as:
\begin{equation}
I(x, y) = \int \rho (x, y, v_{\rm los})dv_{\rm los},
\end{equation}
where $\rho(x, y, v_{\rm los})$ represents the intensity within the PPV cube, and $v_{\rm los}$ is the LOS velocity. $\rho(x, y, v_{\rm los})$ often corresponds to brightness or antenna temperature in observation.

Velocity centroids map, $C(x, y)$, provides an intensity-weighted average velocity for each cell in the map, integrated along the LOS. They are calculated by:
\begin{equation}
C(x, y) = \frac{\int \rho (x, y, v_{\rm los})v_{\rm los}dv_{\rm los}}{\int \rho (x, y, v_{\rm los})dv_{\rm los}}.
\end{equation}

Velocity channel maps, $p(x, y)$, are derived by integrating the PPV cube over a narrowly defined velocity range. These channel maps are predominantly affected by velocity fluctuations due to the phenomenon known as velocity caustics \citep{lazarian2000velocity, 2023MNRAS.524.2379H}. The expression for a thin velocity channel is given by:
\begin{equation}
p(x, y) = \int^{v_0+\Delta v/2}_{v_0-\Delta v/2} \rho (x, y, v_{\rm los})dv_{\rm los},
\end{equation}
where $v_0$ is the center velocity of the channel and $\Delta v$ is the channel width, which is chosen to be less than the square root of the turbulence velocity dispersion, i.e., $\Delta v < \sqrt{(\delta v)^2}$.

These three maps contain complementary information on the density and velocity fields.
The intensity map primarily traces the integrated density along the line of sight, and therefore carries information mainly about the underlying density structure. 
In contrast, the velocity centroid map incorporates both density and velocity information. It provides an intensity-weighted measure of the line-of-sight velocity; thus, it is sensitive to the velocity field when the density fluctuations are moderate. The velocity channel maps, on the other hand, are dominated by velocity fluctuations due to the velocity caustics effects \citep{lazarian2000velocity, 2023MNRAS.524.2379H}. In general, the three selected feature maps characterize different proportions of the density and velocity field information.

\subsection{Neural Net Architecture}
We used a deep learning neural net to infer the mode energy partition from morphological features.
The model employs a Conditional Residual Neural Network following the architecture outlined by \citet{2025ApJ...990...76H}, which was designed to predict 3D magnetic fields from spectroscopic observation. The model integrates conditional information into the prediction process through Feature-wise Linear Modulation (FiLM; \cite{perez2018film}).The model includes three primary characteristics:
(1) Condition Embedding Network \citep{kang2018conditional}, which transforms input conditioning information into a conditioning vector. (2) Encoder-Decoder Structure \citep{aitken2021understanding}, which processes spectroscopic images with instance normalization and reflection padding, followed by downsampling and subsequent upsampling via transposed convolutions. (3) Conditional Residual Blocks \citep{he2016deep}: Each residual block incorporates FiLM conditioning, dynamically modulating intermediate feature maps based on the embedded conditioning vector. The resulting model structure could effectively capture spatial features at multiple scales and in each Conditional Residual Block, which allows the network to adapt its internal representations to varying input conditions.

\subsection{Training Strategy}
The training and evaluation datasets consist of paired observational velocity images (Channel, Centroid, Intensity) and three corresponding output energy partition maps (Alfv\'en, Fast, Slow). 
80\% of the dataset is used in training, and the remaining 20\% is utilized as a validation dataset.
All input spectroscopic images are normalized.
We implemented rotational variations and Gaussian blurring to simulate observational uncertainties and instrumental effects, respectively.
The model is trained with $5,000$ epochs using the mean squared error loss function, optimized via the Adam optimizer with a learning rate of $2\times 10^{-4}$ and momentum parameters (betas) set to 0.5 and 0.999. 
To mitigate overfitting, we monitor the validation loss for early stopping and evaluate the peak signal-to-noise ratio (PSNR) on a held-out test set every 50 epochs. 
The final model corresponds to the checkpoint with the highest PSNR on untransformed test data. 

\begin{figure*}[htbp!]
    \centering
    \includegraphics[width=0.7\linewidth]{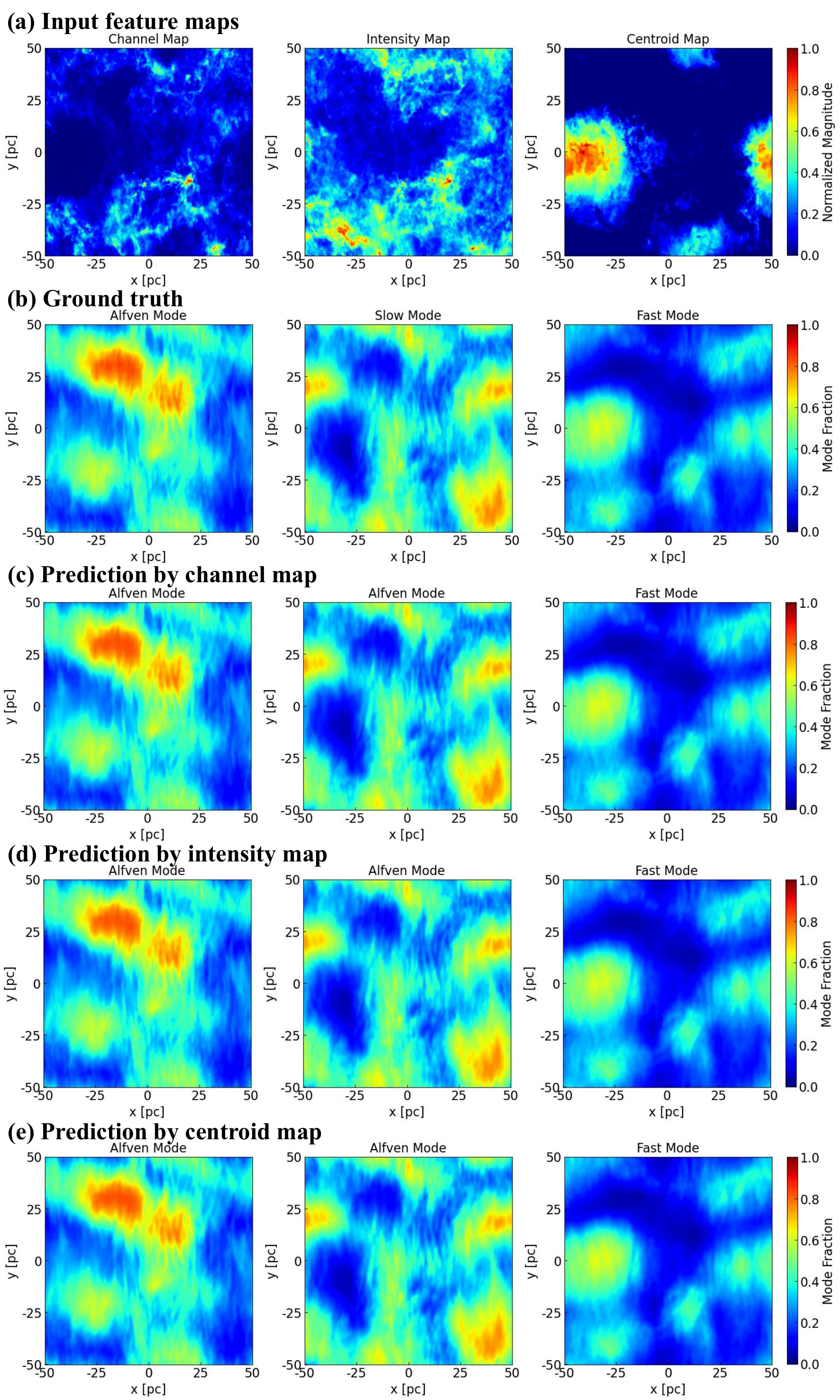}
    \caption{A comparison of the predicted mode fraction and the ground truth using simulation B3v2.5. Panel (a): the input feature maps. Panel (b): the ground truth. Panel (c)-(e): the predicted mode fraction with different feature maps as training inputs.}
    \label{fig:prediction_results}
\end{figure*}


\begin{figure*}{}
    \centering
    \includegraphics[width=0.9\linewidth]{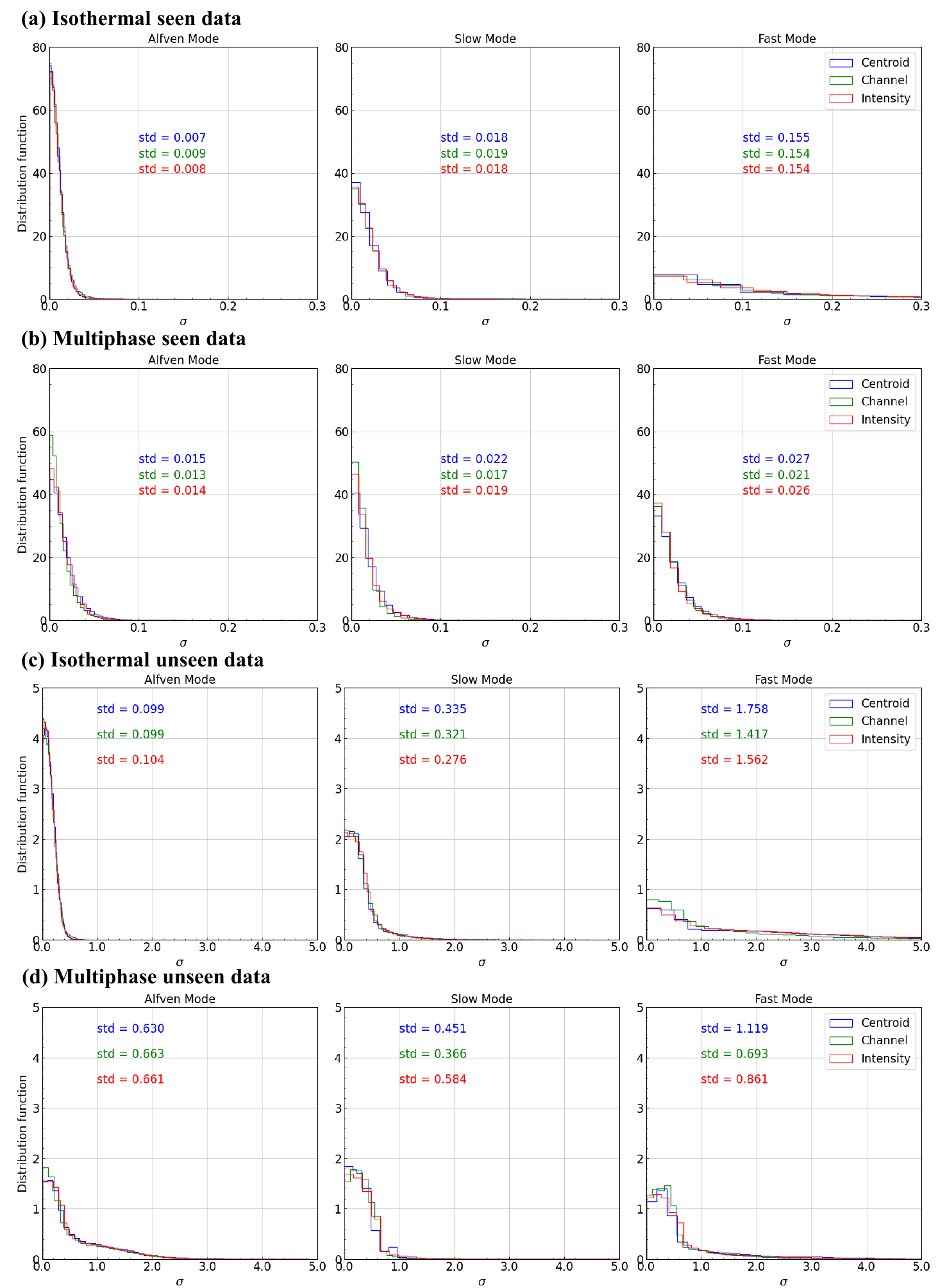}
    \caption{Histograms of the relative prediction error $\sigma$, which is defined as the difference between the predicted ($f^{\rm pred}$) and true $f^{\rm ture}$ energy fractions, normalized by the ground-truth value for each mode, i.e., $\sigma = |f_i^{\rm pred}-f_i^{\rm true}|/f_i^{\rm true}$, where the subscript $i = A, S, {\rm or}~F$ denotes the Alfv\'en, slow, and fast modes, respectively. Here we demonstrate simulation min01 in the isothermal case and simulation B3v2.5 in the multiphase case. Panel (a)-(b): $\sigma$ distribution when data are included in the training inputs. Panel (c)-(d): $\sigma$ distributions when data are excluded in the training inputs. Each panel, from left to right, represents Alfv\'en, slow, and fast modes, respectively. Blue, green, and red refer to different training inputs. "std" means the standard deviation.}
    \label{fig:sigma_distribution}
\end{figure*}

\section{Results}
\label{sec:results}
\subsection{Mode Decomposition in Multi-phase ISM Simulations}
Fig.~\ref{beta_modeenergy} shows the kinetic‐energy fractions of the three MHD modes as functions of plasma $\beta$ in the multi-phase ISM simulations, compared with those in isothermal simulations. The energy fraction of each mode is defined as the ratio between its kinetic energy and the sum of kinetic energies over all MHD modes, i.e. $f_i = E_i/(E_A + E_f + E_s)$, where $E_i = 0.5v_i^2$ and $v_i$ is the velocity of the corresponding mode.
We find that the Alfv\'en mode consistently dominates, carrying approximately 40–65\% of the total kinetic energy as $\beta$ increases from $10^{-3}$ to $10^{-1}$. The fast mode contributes the smallest fraction, rising from about 5-25\% over the same $\beta$ range. The slow-mode fraction remains intermediate between the two. Relative to isothermal conditions, multiphase simulations exhibit an enhanced fast-mode energy fraction of about 20 - 30\%.

In Fig.~\ref{fig:mode_precentage}, we further divide the medium into the cold neutral medium (CNM; $T < 200$~K), unstable neutral medium (UNM; $200~\mathrm{K} < T < 5000~\mathrm{K}$), and warm neutral medium (WNM; $T > 5000$~K), and repeat the analysis of the kinetic‐energy fractions of the three MHD modes. In the CNM, the energy fractions of the Alfv\'en and slow modes are comparable, each contributing about 30–40\%. The fraction of the Alfv\'en mode decreases with increasing $\beta$, while that of the slow mode increases. Compared with the isothermal case at a similar $\beta$, the Alfv\'en‐mode fraction in the CNM is smaller, whereas the fast‐mode fraction is larger, reaching roughly 30\%. The energy‐partition trends in the UNM and WNM are qualitatively similar to those in the CNM.

The absence of thermal pressure feedback in isothermal simulations primarily causes this difference. In multi-phase ISM, thermal pressure variations strongly couple with density fluctuations, enabling efficient energy transfer between fast and slow modes through compression and rarefaction of gas at varying temperatures. In contrast, isothermal conditions suppress this thermal coupling, forcing the turbulent cascade to redistribute energy differently among the MHD modes, thereby favoring Alfv\'en modes over thermally mediated compressive modes.

\subsection{Comparison of Model Prediction and Ground Truth}
Fig.~\ref{fig:prediction_results} presents a representative case from the simulation with $B \sim 3~\SI{}{\micro G}$ and $\sigma_v \sim 2.5~{\rm km~s^{-1}}$, illustrating how distinct local structures in the feature maps arise from different fractions of the three MHD modes. An example of compressively driven turbulence is presented in Appendix's Fig.~\ref{fig:prediction_results_compressible}. The channel, intensity, and centroid maps each display characteristic morphological signatures. In regions where the fast mode dominates—specifically the left–middle and right–middle portions of the map—the centroid field exhibits two prominent, high-amplitude isotropic structures, whereas the corresponding channel map shows two low-intensity features. These features reflect the intrinsically isotropic and compressive nature of the fast mode. 

We also compare the ResNet prediction with the ground-truth mode energy fractions in Fig.~\ref{fig:prediction_results}. The model accurately recovers the spatial distribution and relative fraction of the three modes. Moreover, the choice of input—channel, intensity, or centroid map has only a marginal impact on the reconstruction accuracy.

Next, we evaluate the prediction performance of the proposed conditional ResNet model in both isothermal and multiphase simulations. Fig.~\ref{fig:sigma_distribution} shows the histograms of the relative prediction error, $\sigma$, defined as the difference between the predicted ($f^{\rm pred}$) and true ($f^{\rm ture}$) energy fractions, normalized by the ground-truth value for each mode, i.e., $\sigma = |f_i^{\rm pred}-f_i^{\rm true}|/f_i^{\rm true}$, where the subscript $i = A, S, {\rm or}~F$ denotes the Alfv\'en, slow, and fast modes, respectively. We consider two evaluation settings: (1) training on all available data and testing on the same (seen) patches, and (2) withholding one patch during training and testing solely on that unseen patch.

For seen data, $\sigma$ remains below 0.3 for all three modes in both isothermal and multiphase simulations, with the fast-mode prediction achieving the highest accuracy (Figs.~\ref{fig:sigma_distribution}(a) and (b)). For unseen data, $\sigma$ generally stays within 2 in both cases, though the distributions develop broader tails. The mean error in the isothermal case ($\sigma \sim 0.10–1.79$) is higher than in the multiphase case ($\sigma \sim 0.35–0.83$), indicating that the model performs better when predicting mode fractions in multiphase turbulence. Alfv\'en-mode predictions consistently outperform those of the fast and slow modes across all scenarios (Figs.~\ref{fig:sigma_distribution}(c) and (d)). Overall, the relative error distributions generally peak at zero across all configurations, except for fast-mode predictions on unseen data, indicating good consistency and predictive robustness of the model. Among the three modes, the Alfv\'en mode exhibits the smallest errors, whereas the fast mode remains the most unstable. This behavior may be partially attributed to the isotropic nature of the fast mode, which leads to less distinct morphological structures in the sliced feature maps.

\section{Discussion}
\label{sec:dis}
\subsection{Comparison with earlier work}
There have been several efforts to identify the dominant MHD modes in the solar system and in diverse Galactic environments. These studies rely primarily on synchrotron polarization analyses—for example, examining $\gamma$-ray emission excesses and the anisotropies encoded in combinations of Stokes parameters \citep{2020NatAs...4.1001Z, 2023MNRAS.524.6102M}. In parallel, \citet{2024MNRAS.52711240H, 2024ApJ...975...66H} introduced convolutional neural network (CNN)–based models capable of capturing anisotropies in spectroscopic and synchrotron observations, enabling reconstruction of the 3D magnetic-field structure and the magnetization level of the medium. 

Building on this foundation, our work explores a new direction by directly targeting the three fundamental MHD modes—Alfv\'en, slow, and fast. We proposed a machine learning model - conditional ReNet - trained on both isothermal and multiphase MHD simulations to infer the mode energy fractions directly from spectroscopic observables. We demonstrate that machine learning can recover the energy distribution among the three modes across a wide range of physical conditions, providing a new avenue for diagnosing turbulence in the interstellar medium.

Furthermore, recent studies have shown that 21 cm H I thin-channel maps, when combined with the Galactic rotation curve, can be used to reconstruct the 3D distribution of the Galactic magnetic field \citep{2023MNRAS.524.2379H, 2024MNRAS.528.3897S, 2025ApJ...990...76H}. A similar idea could be incorporated into our ML framework, potentially enabling the reconstruction of the 3D distribution of turbulence-mode energy fractions throughout the ISM.

\subsection{Implications for studying cosmic ray transport and star formation}
Our ability to characterize the turbulence‐mode composition in both isothermal and multi-phase media has important implications for understanding cosmic-ray (CR) transport. Fast modes are particularly effective at pitch-angle scattering and stochastic acceleration because of their compressive and nearly isotropic nature \citep{2002PhRvL..89B1102Y, 2008ApJ...673..942Y, 2013ApJ...779..140X}. Our finding that low-$\beta$ ($\sim 0.1$) multi-phase media exhibit enhanced fast-mode fractions compared to the isothermal case suggests that CR scattering may be correspondingly stronger in such environments. This enhancement directly affects the diffusion and acceleration of CRs in the ISM.

Furthermore, turbulence plays a crucial role in star formation. Different fraction of compressive components leads to different star formation rates \citep{2008ApJ...688L..79F,2012ApJ...761..156F}. The ML-assisted framework developed in this work provides a new avenue for observationally constraining turbulence-mode composition using spectroscopic observations. This, in turn, enables more accurate modeling of CR propagation and star formation in the multi-phase ISM.

\subsection{Limitations}
Several limitations merit consideration. First, both the training and validation datasets are derived from numerical simulations and are therefore subject to numerical artifacts. In particular, simulations inevitably introduce numerical dissipation of turbulence, whereas such dissipation is not present in real observations. Second, observational data are affected by instrumental noise and finite beam size, neither of which is included in the current simulations. We do not expect noise to significantly degrade the model performance: in our earlier studies on predicting the sonic Mach number and magnetic field strength \citep{2025ApJ...982..121S,2025ApJ...990...76H}, the machine-learning framework remained robust for signal-to-noise ratios exceeding three.

In addition, the experiments presented here explore only a restricted region of parameter space. All simulations are performed in the strong-field, sub-Alfv\'enic ($M_A \le 1$) regime, whereas the interstellar medium may be super-Alfv\'enic ($M_A > 1$) on large scales. Although the ResNet exhibits robust performance across both isothermal and multiphase conditions, its ability to generalize to super-Alfv\'enic turbulence has yet to be assessed.

\section{Conclusion}
\label{sec:con}
In this study, we investigated the kinetic-energy fractions of the three fundamental MHD turbulence modes, Alfv\'en, slow, and fast modes, under both isothermal and multiphase ISM conditions. We further developed a method that employs a conditional Residual Neural Network (ResNet) together with three distinct spectroscopic maps to estimate these mode fractions. This approach leverages the morphological differences imposed by distinct MHD modes and presents a novel method for determining mode composition from spectroscopic observations. Our main findings are as follows:
\begin{enumerate}
\item The Alfv\'en and slow modes carry comparable kinetic-energy fractions and together dominate the turbulent energy budget in multiphase media, while the fast mode contributes the smallest fraction. Relative to the isothermal case, multiphase simulations show a modest reduction in Alfv\'en-mode energy fraction and a corresponding enhancement in the fast-mode energy fraction.
\item We implemented a ResNet capable of inferring mode energy fractions from three observational feature maps—channel maps, centroid velocity maps, and integrated-intensity maps—each encoding different combinations of velocity and density information.
\item Training the model on both isothermal and multiphase simulations shows that fast-mode predictions using channel maps achieve the lowest errors. The mean relative normalized errors of approximately 0 and the standard deviation of 0.01 - 0.02 for seen data and 0.1 - 1.8 for unseen data.
\end{enumerate}

\begin{acknowledgments}
Y.H. acknowledges the support for this work provided by NASA through the NASA Hubble Fellowship grant No. HST-HF2-51557.001 awarded by the Space Telescope Science Institute, which is operated by the Association of Universities for Research in Astronomy, Incorporated, under NASA contract NAS5-26555. This work used SDSC Expanse CPU/GPU and NCSA Delta CPU/GPU through allocations PHY230032, PHY230033, PHY230091, PHY230105,  PHY230178, and PHY240183, from the Advanced Cyberinfrastructure Coordination Ecosystem: Services \& Support (ACCESS) program, which is supported by National Science Foundation grants \#2138259, \#2138286, \#2138307, \#2137603, and \#2138296. 
\end{acknowledgments}

\newpage
\appendix
\section{Compressive multiphase simulation}
\label{appendix}

\begin{figure*}{}
    \centering
    \includegraphics[width=0.9\linewidth]{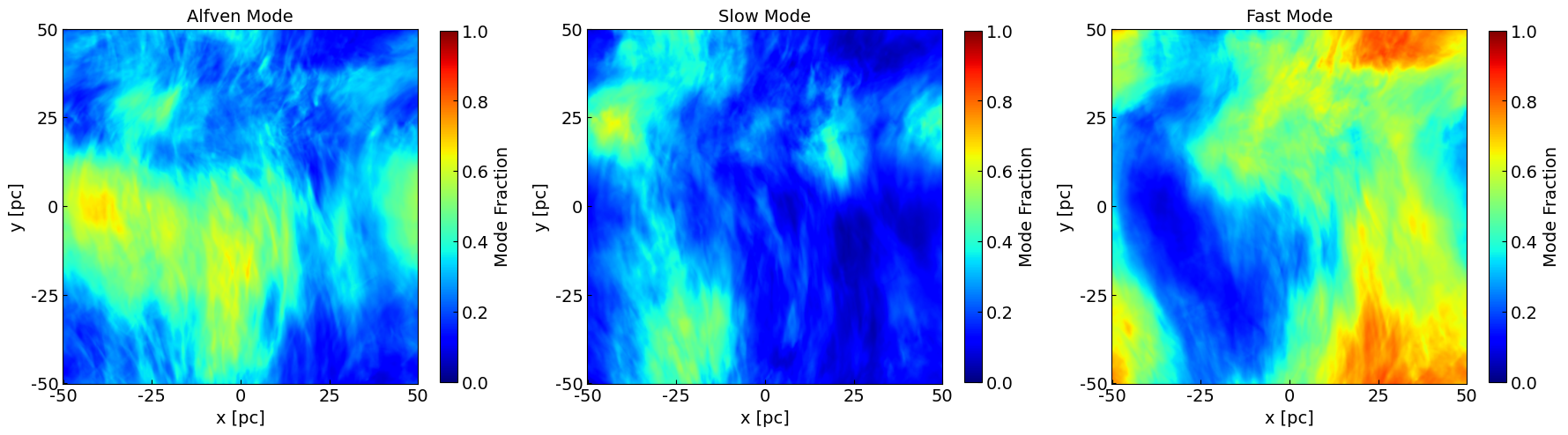}
    \caption{Illustration of three MHD turbulence modes' kinetic energy fraction for a compressively driven turbulence simulation after mode decomposition.  The simulation is conducted by initial setup $B \approx 5~\SI{}{\micro G}$, $\sigma_v = 10~\text{km s}^{-1}$ ($\beta = 0.2015$).}
    \label{fig:mode_illustration_compressible}
\end{figure*}

\begin{figure*}{}
    \centering
    \includegraphics[width=0.75\linewidth]{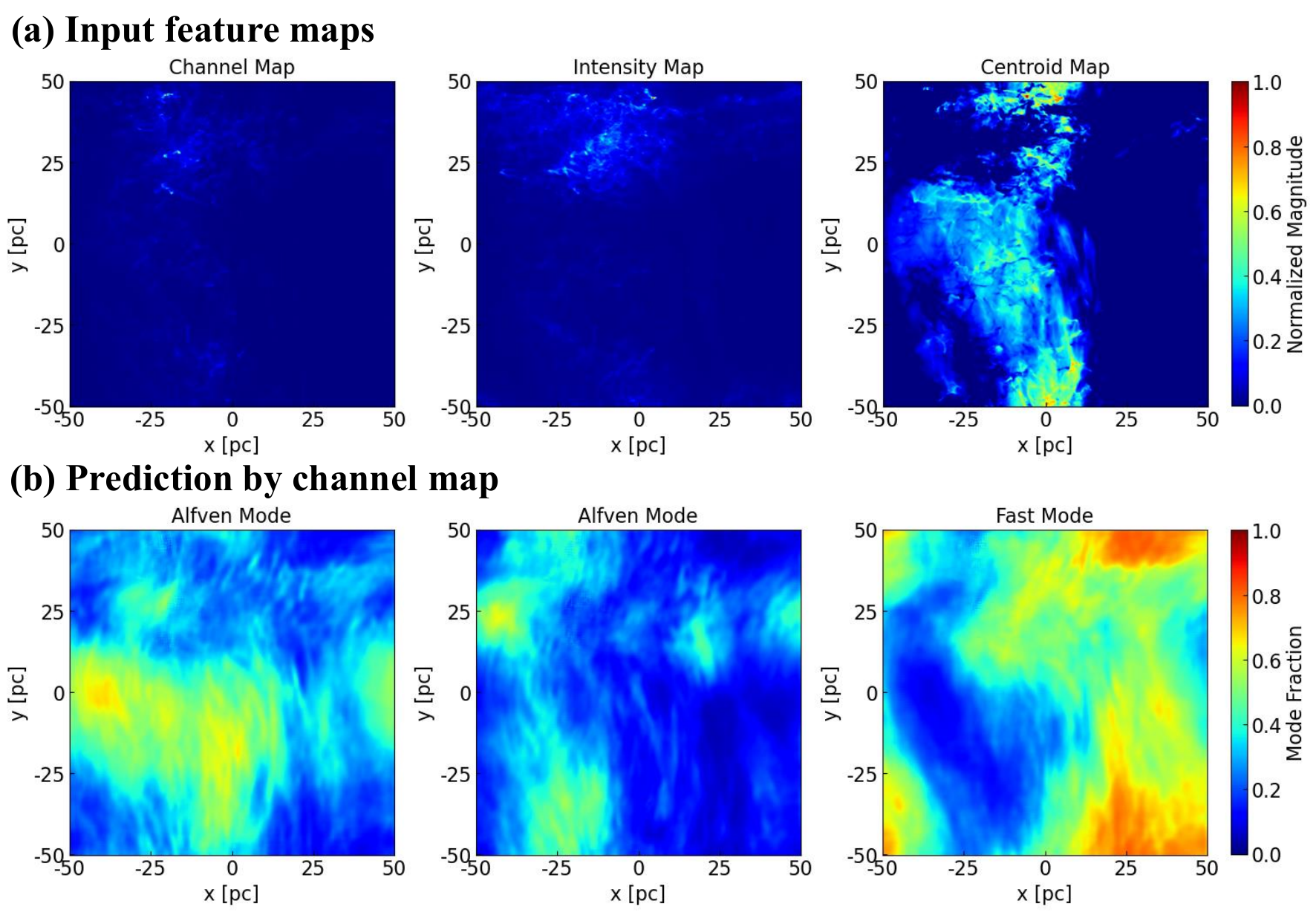}
    \caption{A comparison of the channel map predicted mode fraction and the ground truth using compressible MHD simulation. Panel (a): the input feature maps. Panel (b): the predicted mode fraction with channel map as training inputs. The ground truth of the prediction is referred to as Fig.~\ref{fig:mode_illustration_compressible}. This simulation is conducted by initial setup $B \approx 5~\SI{}{\micro G}$, $\sigma_v = 10~\text{km s}^{-1}$ ($\beta = 0.2015$).}
    \label{fig:prediction_results_compressible}
\end{figure*}

We additionally performed mode decomposition for a multiphase simulation with fully compressively driven turbulence (Fig.~\ref{fig:mode_illustration_compressible}). This simulation adopts an initial magnetic field strength of $B \approx 5~\SI{}{\micro G}$ and a velocity dispersion of $\sigma_v = 10~\mathrm{km,s^{-1}}$, corresponding to an average plasma beta of $\beta = 0.2015$. Compared to the solenoidally driven case in \S~\ref{sec:results}, the fraction of fast-mode energy increases substantially, reaching $\sim 0.2 - 0.8$ in the compressible simulation, consistent with \cite{2020PhRvX..10c1021M}.

We further evaluated the ResNet performance in estimating mode energy fractions using this same compressible simulation (Fig.~\ref{fig:prediction_results_compressible}). The predicted mode fractions recover the ground-truth values. This behavior persists across a broader set of input channel maps, indicating that the ResNet architecture generalizes well to both solenoidally driven and compressively driven MHD turbulence when estimating mode energy fractions.



\newpage
\bibliography{sample631}{}
\bibliographystyle{aasjournal}



\end{document}